\documentclass[twocolumn,prb,notitlepage,floats,superscriptaddress,amsmath,amssymb]{revtex4-2}

\usepackage[version=3]{mhchem} 
\usepackage{bm}

\usepackage[utf8]{inputenc}
\usepackage[T1]{fontenc}
\usepackage{graphicx}
\usepackage{upgreek}
\usepackage{color}
\usepackage{ulem}
\usepackage{hyperref} 
\hypersetup{colorlinks,citecolor=blue, filecolor=blue ,linkcolor=blue , urlcolor=blue, pdftex}
\usepackage{sidecap}
\usepackage{amssymb}
\usepackage[caption=false]{subfig}

\def \FUW{Faculty of Physics, University of Warsaw, 02-093 Warsaw, Poland}
\def \Wroclaw{Institute of Theoretical Physics, Wrocław University of Science and Technology, 50-370 Wrocław, Poland}
\def \Shenzhen{Guangdong Provincial Key Laboratory of Nano-Micro Materials Research, School of Advanced Materials,  Peking University Shenzhen Graduate School, Shenzhen, 518055, China}
\def \LNCMI{LNCMI-EMFL, CNRS, UPR 3228, Universit´e Grenoble Alpes, 38000 Grenoble, France}

\begin{document}
	
	\title{Resonant Raman scattering in bilayer 3R-MoS$_{2}$ }
	
	\author{Chinmay K. Mohanty}
	\email{c.mohanty@uw.edu.pl}
	\affiliation{\FUW}
	\author{Kacper Walczyk}
	\affiliation{\FUW}
	\author{Tomasz Wo\'zniak}
	\affiliation{\Wroclaw}
	\author{Chengcheng~Jiang}
	\affiliation{\Shenzhen}
	\author{Adam Babi\'nski}
	\affiliation{\FUW}
	\author{Clement Faugeras}
	\affiliation{\LNCMI}
	\author{Zhaolong Chen}
	\affiliation{\Shenzhen}
	\author{Maciej R. Molas}
	\email{maciej.molas@fuw.edu.pl}
	\affiliation{\FUW}

	\begin{abstract}
		Raman scattering is a powerful spectroscopic technique widely employed to investigate light–matter interactions and lattice dynamics in two-dimensional materials. 
		Here, we investigate the temperature-dependent resonant Raman response of bilayer 3R-MoS$_2$. 
		The study combines multi-wavelength Raman spectroscopy, photoluminescence measurements, and density functional theory calculations to track the evolution of excitonic transitions and resonance conditions. 
		We observe contributions from both zone-centre and finite-momentum phonons, a pronounced quenching of the Stokes intensity at low temperatures followed by saturation, the emergence of anti-Stokes scattering above 130~K, and a strong deviation of the effective phonon temperature from the lattice temperature induced by resonance effects. 
		These results demonstrate that the Raman response is governed by the interplay between incoming and outgoing resonance processes, providing deeper insight into exciton--phonon coupling in van der Waals materials.
		
	\end{abstract}
	
	\maketitle
	
	\section*{Introduction \label{sec:Intro}}
	Two-dimensional (2D) van der Waals (vdW) materials have attracted considerable attention owing to their unique electronic properties and potential for optoelectronic applications~\cite{Lopez-Sanchez2013, Jo2014, Mak2016, Zhu2016, Yan2017, wang2026all, kim2025review, thayil2026nanostructured, javeed2025recent}.
	Such materials exhibit strong quantum confinement and reduced dielectric screening, resulting in enhanced Coulomb interactions and pronounced excitonic effects~\cite{huser2013dielectric,chhowalla2013chemistry,molas2019probing, slobodeniuk2023exciton}. 
	Among these, transition metal dichalcogenides (TMDs) constitute an important family of semiconducting 2D materials, due to their thickness-dependent band structure and rich optical response~\cite{mak2010atomically,wang2012electronics,jin2013direct}. 
	In particular, TMDs undergo a transition from an indirect band gap in the bulk to a direct band gap in the monolayer limit, leading to strong photoluminescence (PL) and rendering them promising candidates for nanoscale optoelectronic and valleytronic devices~\cite{ mak2010atomically, jiang2014valley,Molas2017,duan2024introduction,ahmed2024retracted,xie2025plasma}.
	
	In multilayer TMDs, the stacking configuration plays a crucial role in determining both vibrational and electronic properties~\cite{Lee2010, Tonndorf2013, Placidi2015, Grzeszczyk2016, Zhang2018, kipczak2020optical}.
	Semiconducting TMD in trigonal prismatic coordination can adopt two principal configurations , namely 2H and 3R, which differ in their symmetry and interlayer coupling.
	In the 2H phase, adjacent layers are rotated by 180$^\circ$ with respect to each other, leading to alternating layer orientations and resulting in centrosymmetric stacking in the bulk.
	In contrast, the 3R phase is characterised by a unidirectional stacking in which successive layers are laterally shifted relative to one another without rotation. 
	This gives rise to rhombohedral symmetry and the absence of inversion symmetry.
	Consequently, 3R-stacked TMDs often display distinct electronic and optical properties~\cite{strachan20213r,Lee2015,deb2024excitonic,aggarwal2025stacking} including a modified band structure and enhanced nonlinear optical response, compared with their 2H counterparts~\cite{golasa2014resonant,latzke2015electronic,bhatnagar2022temperature}.
	
	Raman scattering (RS) is a powerful technique to probe lattice vibrations. Under resonant conditions, when the excitation energy approaches excitonic transitions, resonance-enhanced RS mediated by exciton--phonon coupling leads to a strong enhancement of the Raman intensity together with the activation of additional phonon modes beyond first-order processes due to the relaxation of selection rules~\cite{golasa2014resonant,Lee2015,Chow2017,Shree2018,kumar2021davydov,bhatnagar2022temperature}. 
	This includes higher-order features and phonons originating from finite-momentum states~\cite{golasa2014resonant,Lee2015,kumar2021davydov,bhatnagar2022temperature}, making resonant Raman scattering a sensitive probe of exciton--phonon interactions.
	The resonance condition can be tuned by varying the excitation energy using multi-wavelength lasers, as well as by temperature, which modifies both the exciton energies and the phonon population~\cite{maher2006temperature, Molas2017, bhatnagar2022temperature, zinkiewicz2024raman}. 
	As a result, the Raman intensity and the relative contribution of different phonon modes, including the balance between Stokes and anti-Stokes processes, are strongly affected, providing direct insight into exciton--phonon coupling. 
	Despite these well-established aspects, the temperature-dependent Raman response under resonant conditions in 3R-MoS$_2$ BL, including the correlation between excitonic resonance and the evolution of Raman intensity as well as the interplay between Stokes and anti-Stokes processes, remains unexplored.
	
	\begin{figure*}[t]
		\subfloat{}%
		\centering
		\includegraphics[width=1\linewidth]{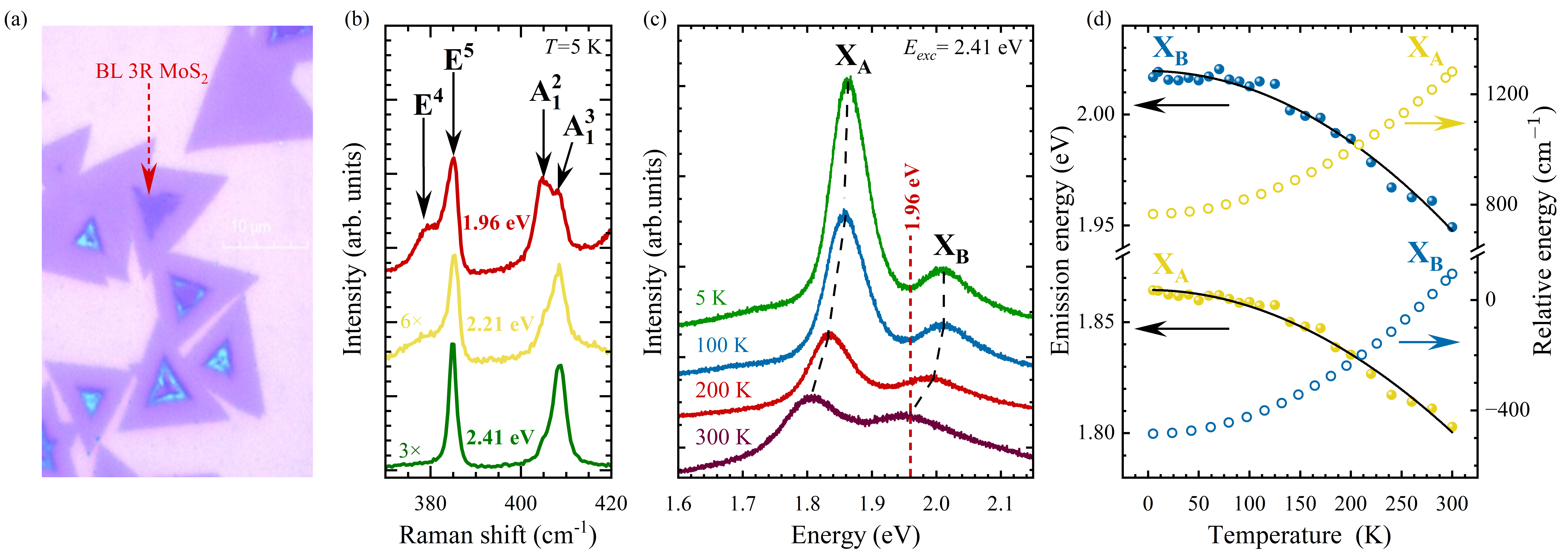}
		\caption{(a) Optical microscopy image of a CVD-grown bilayer 3R-MoS$_2$ flake, indicated by a red arrow.
			(b) Raman scattering spectra of bilayer 3R-MoS$_2$ measured at 5~K using excitation energies of 1.96, 2.21, and 2.41~eV with an excitation power of 300~$\mu$W.
			(c) Temperature evolution of PL spectra obtained under 2.41~eV laser excitation. The black dashed lines denote the energies of the $\mathrm{X_A}$ and $\mathrm{X_B}$ lines. 
			The vertical red dashed line indicates the energy of the 1.96~eV excitation.
			(d) Emission energies of $\mathrm{X_A}$ (yellow) and $\mathrm{X_B}$ (blue) as a function of temperature. 
			Solid black curves represent Varshni fits, and open yellow and blue symbols correspond to the energies of $\mathrm{X_A}$ and $\mathrm{X_B}$ relative to the excitation energy (1.96~eV), respectively.}
		\label{fig:fig1}
	\end{figure*}
	
	
	In this work, we investigate the temperature-dependent resonant Raman scattering in bilayer 3R-MoS$_2$, focusing on the interplay between excitonic resonances and lattice dynamics. 
	By combining Raman spectroscopy performed under multi-wavelength excitation with temperature-dependent photoluminescence measurements, we track the evolution of excitonic transitions and their influence on the vibrational response.
	Using the Varshni model, we quantify the temperature dependence of the A and B excitons and demonstrate a continuous tuning of the resonance conditions over a wide temperature range. We show that, under resonant excitation, the RS response involves contributions from both zone-centre and finite-momentum phonons, including multiphonon processes activated by exciton–phonon coupling.
	Furthermore, by analysing the temperature evolution of the Stokes and anti-Stokes scattering intensities and the effective phonon temperature, we reveal the dominant role of resonant Raman scattering mechanisms. 
	In particular, we identify the competition between incoming and outgoing resonance processes associated with the B exciton as the key factor governing the observed enhancement of Raman modes.
	
	\begin{figure*}[t]
		\centering
		\includegraphics[width=1\linewidth]{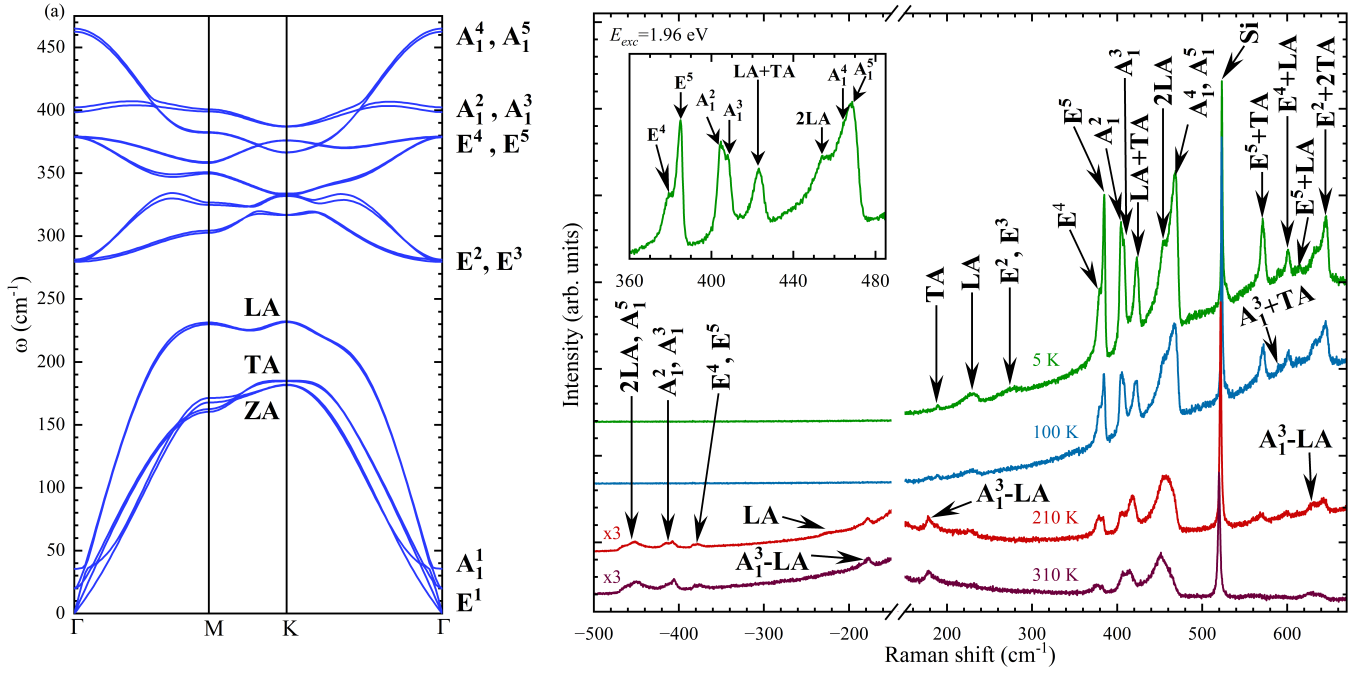}
		\caption{(a) Calculated phonon dispersion of 3R-stacked MoS$_{2}$ BL. Raman-active modes at the $\Gamma$ point of the BZ are indicated.
			(b) Temperature evolution of the Stokes and anti-Stokes branches in high-resolution RS spectra measured under resonant excitation (1.96 eV) at $T$ = 5, 100, 210, and 310 K.
			The assigned phonon peaks, including Raman-active and acoustic modes, as well as their combinations arising from higher-order processes, are labelled.  
		}
		\label{fig:fig2}
	\end{figure*}
	
	\section*{Results and Discussion \label{results}}
	\subsection*{Optical and vibrational properties of bilayer 3R-MoS$_2$}
	
	The CVD technique was employed to grow bilayer 3R-MoS$_{2}$ crystals on a SiO$_{2}$/Si substrate (see Methods for details), as shown in Fig.~\ref{fig:fig1}(a). 
	Optical microscopy reveals triangular domains with well-defined edges, characteristic of high-quality MoS$_{2}$ flakes~\cite{najmaei2013vapour,aggarwal2025stacking}. 
	The 3R phase of the investigated 3R-MoS$_2$ BL was confirmed using second harmonic generation (SHG) measurements; see the Supplementary Information (SI) for details. 
	The non-centrosymmetric nature of the 3R phase results in a strong SHG signal, whereas SHG is suppressed in 2H-stacked BLs due to inversion symmetry~\cite{kumar2013second, li2013probing}.
	
	The crystal structure of BL 3R-MoS$_2$ belongs to the space group  P3m1 (No.~156), corresponding to the point group $\mathrm{C_{3v}}$~\cite{coutinho20173r, liang2022optically, strachan20213r}. 
	The normal modes of lattice vibrations at the $\Gamma$ point of the Brillouin zone (BZ) can be expressed in terms of the following irreducible representations: $\Gamma = 6\mathrm{A}_1 + 6\mathrm{E}$.
	Restricting to optical phonon modes and excluding the acoustic modes $\mathrm{A}_1$ (LA/ZA) and $\mathrm{E}$ (TA), the lattice vibrations at the $\Gamma$ point are given by: $\Gamma_{\mathrm{optical}} = 5\mathrm{A}_1 + 5\mathrm{E}$.
	
	To investigate the influence of excitation energy on the RS response, low-temperature ($T = 5$~K) Raman spectra were measured using excitation energies of 2.41, 2.21, and 1.96~eV, as shown in Fig.~\ref{fig:fig1}(b).
	Under 2.41~eV excitation, the RS spectrum is dominated by the characteristic first-order modes, namely $\mathrm{E}^2$ and $\mathrm{A}^3_{1}$~\cite{zheng2017high,zhang2020near,ullah2021selective,aggarwal2025stacking}. 
	Here, the superscripts $1$, $2$, $3$, and so on are used to distinguish between different $\mathrm{A_1}$ and $\mathrm{E}$ modes of the same symmetry.
	The energy separation between these modes is $\sim 23$~cm$^{-1}$, confirming the bilayer nature of the sample, in agreement with previous reports~\cite{zheng2017high,ullah2021selective,xu2024reconfiguring,aggarwal2025stacking}.
	Upon tuning the excitation energy to 2.21~eV and 1.96~eV, additional features emerge in the RS spectra, including modes associated with $\mathrm{E}^{4}$ and $\mathrm{A}^{3}_{1}$, which may originate from Davydov-like splitting induced by interlayer coupling, as previously reported in multilayer TMD systems~\cite{Sekine1984,Na2018, Shinde2021,bhatnagar2022temperature}. 
	Their appearance indicates resonant Raman scattering via exciton--phonon coupling, accompanied by a pronounced increase in the intensity of the $\mathrm{E}^5$ and $\mathrm{A^3_{1}}$ modes, reaching a maximum at 1.96~eV, as discussed below.
	A similar effect has been reported for 2H MoS$_2$ BLs~\cite{golasa2014resonant,Lee2015,bhatnagar2022temperature}.
	
	To reveal the exciton--phonon coupling, we measured the temperature evolution of the PL response, as shown in Fig.~\ref{fig:fig1}(c). 
	At low temperature ($T = 5$~K), the spectrum exhibits two prominent emission peaks corresponding to the neutral excitons, denoted as A ($\mathrm{X_A}$) and B ($\mathrm{X_B}$), originating mainly from the spin--orbit splitting of the valence band at the $K^+$ and $K^-$ valleys of the BZ~\cite{ jiang2014valley,shi20173r,paradisanos2020controlling, grzeszczyk2021optical}. 
	With increasing temperature, a pronounced redshift of the emission energies is observed, accompanied by a reduction in the PL intensity, consistent with bandgap renormalisation and enhanced non-radiative recombination processes. 
	The $\mathrm{X_A}$ exciton remains the dominant feature over the entire temperature range, whereas the $\mathrm{X_B}$ exciton appears as a weaker high-energy peak. 
	Notably, the temperature-induced redshift causes the $\mathrm{X_A}$ transition to move progressively away from the 1.96~eV excitation energy, while the $\mathrm{X_B}$ transition moves closer to it, indicating continuous tuning of the resonance condition between the A and B excitons.
	To quantify the temperature evolution of the excitonic transitions, the emission energies were extracted and plotted as a function of temperature in Fig.~\ref{fig:fig1}(d). 
	Both $\mathrm{X_A}$ and $\mathrm{X_B}$ lines exhibit a monotonic redshift with increasing temperature, which can be well described by the Varshni relation:
	\begin{equation}
		E(T) = E(0) - \frac{\alpha T^2}{T + \beta},  
	\end{equation}
	where $E(0)$ is the transition energy at 0~K, and $\alpha$ and $\beta$ are material-specific fitting parameters describing the strength of electron--phonon coupling and the characteristic temperature scale, respectively. 
	The fitted curves are plotted in Fig.~\ref{fig:fig1}(d), showing excellent agreement with the experimental data.
	The extracted fitting parameters for the $\mathrm{X_A}$ and $\mathrm{X_B}$ transitions are summarised in Table~\ref{tab:varshni}.
	In addition, we evaluated the relative energy separation between the 1.96~eV excitation laser and the excitonic transitions (see Fig.~\ref{fig:fig1}(d)).
	The energy difference between the 1.96~eV excitation and the $\mathrm{X_A}$ emission increases from approximately 800~cm$^{-1}$ at low temperature to nearly 1300~cm$^{-1}$ at 300~K, while the corresponding separation for the $\mathrm{X_B}$ transition evolves from about $-500$~cm$^{-1}$ to approximately 100~cm$^{-1}$.
	This confirms that the resonant conditions for RS in 3R MoS$_2$ BL can be continuously tuned between the two excitonic resonances over the 5--300~K temperature range. 
	This behaviour plays a crucial role in governing the temperature-dependent Raman response discussed in the following sections.

	\begin{table}[h]
		\centering
		\caption{Fitted Varshni parameters for the $\mathrm{X_A}$ and $\mathrm{X_B}$ excitonic transitions.}
		\label{tab:varshni}
		\begin{tabular*}{\linewidth}{@{\extracolsep{\fill}}c||c|c|c}
			Exciton & E(0) (eV) & $\alpha$ (meV/K) & $\beta$ (K) \\
			\hline\hline
			$\mathrm{X_A}$ & 1.865 & 0.552 & 505 \\
			$\mathrm{X_B}$ & 2.021 & 0.626 & 512 \\
		\end{tabular*}
	\end{table}

	\begin{figure*}[!t]
		\subfloat{}%
		\centering
		\includegraphics[width=1\linewidth]{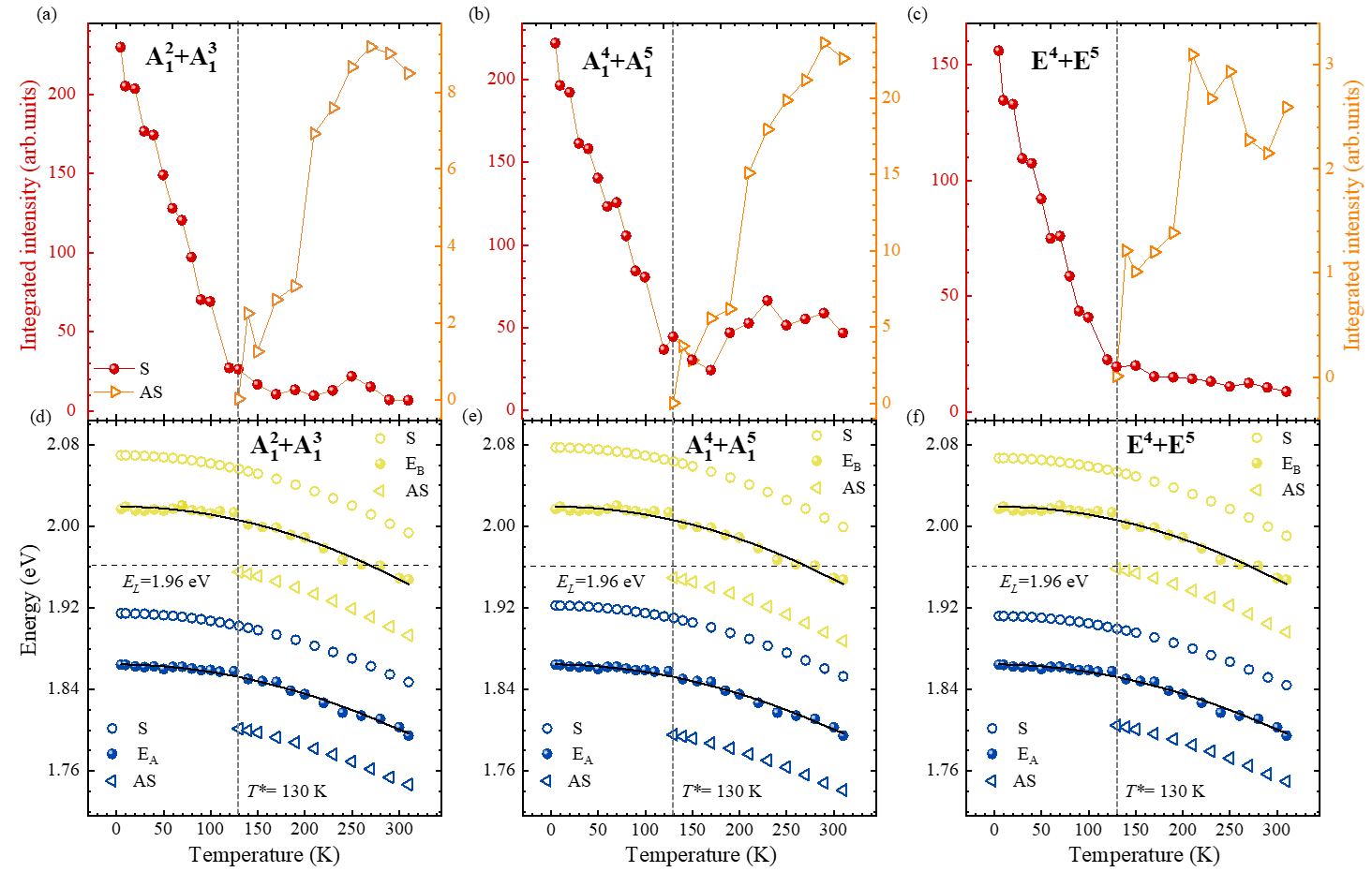}
		\caption{Temperature evolution of the resonant Raman response of bilayer 3R-MoS$_2$. Panels (a)--(c) show the integrated intensities of the Stokes (red points) and anti-Stokes (orange points) components for the $\mathrm{A}_1^{2}+\mathrm{A}_1^{3}$, $\mathrm{A}_1^{4}+\mathrm{A}_1^{5}$, and $\mathrm{E}^{4}+\mathrm{E}^{5}$ modes, respectively. Note that the vertical axis ranges are adjusted to the maximum intensity of each phonon mode. Panels (d)--(f) present the corresponding Stokes and anti-Stokes scattering energies together with the $\mathrm{X_A}$ and $\mathrm{X_B}$ excitonic transitions. The horizontal dashed line denotes the excitation energy $E_{\mathrm{L}} = 1.96$~eV, while the solid black curves represent the Varshni fits of the excitonic transitions. The vertical dashed line marks the characteristic temperature $T^{*}\approx130$~K, above which the anti-Stokes signal becomes experimentally detectable.}  
		\label{fig:fig3}
	\end{figure*}

	\subsection*{Temperature effect on resonant Raman scattering}
	
	The phonon dispersion shown in Fig.~\ref{fig:fig2}(a), calculated within the density functional theory framework, provides a reference for lattice dynamics across the Brillouin zone (BZ) and serves as a basis for assigning the experimentally observed Raman features.
	It highlights both optical and acoustic phonon branches, including modes of $\mathrm{E}$ and $\mathrm{A_1}$ symmetry, as well as low-energy acoustic branches (LA, TA, and ZA).
	The corresponding phonon density of states is provided in Fig.~S3 of the Supplementary Information
	Figure~\ref{fig:fig2}(b) presents the Raman spectra of the investigated BL measured under 1.96~eV excitation at selected temperatures, focusing on both the Stokes and anti-Stokes regimes. 
	The Stokes component reveals phonon modes over the entire temperature range, whereas the anti-Stokes signal becomes detectable only above 130~K and increases progressively with temperature, consistent with the thermal population of phonon states governed by Bose–Einstein statistics~\cite{hart1970temperature,menendez1984temperature}. 
	The Raman spectrum measured at 5~K consists of eight phonon modes associated with intralayer atomic vibrations, namely $\mathrm{E}^2$ (277~cm$^{-1}$), $\mathrm{E}^3$ (279~cm$^{-1}$), $\mathrm{E}^4$ (378~cm$^{-1}$), $\mathrm{E}^5$ (382~cm$^{-1}$), $\mathrm{A}^2_{1}$ (402~cm$^{-1}$), $\mathrm{A}^3_{1}$ (404~cm$^{-1}$), $\mathrm{A}^4_{1}$ (464~cm$^{-1}$), and $\mathrm{A}^5_{1}$ (466~cm$^{-1}$), in agreement with the calculated phonon dispersion, which predicts eight Raman-active modes of $\mathrm{E}$ and $\mathrm{A_1}$ symmetry at the $\Gamma$ point of the BZ.
	
	Moreover, the 5~K RS response in the low-frequency region below 280~cm$^{-1}$ reveals contributions from phonons with finite momentum. 
	The features observed at 188~cm$^{-1}$ and 226~cm$^{-1}$ can be attributed to acoustic phonons, corresponding to LA and TA modes near the K point of the BZ~\cite{zhang2015phonon,Shinde2021}.
	The activation of these modes reflects a relaxation of momentum conservation under resonant excitation, allowing phonons from across the BZ to contribute to the RS process~\cite{golasa2015disorder, soubelet2016resonance,gontijo2019double,tan2021breakdown}.
	These excitation conditions also result in the appearance of several Raman peaks arising from higher-order, so-called multiphonon processes. 
	We identify Raman features associated with the combination of two acoustic phonons, namely LA+TA (423~cm$^{-1}$), and the double longitudinal acoustic phonon (2LA) mode at 452~cm$^{-1}$.
	In the frequency range from 570 to 650~cm$^{-1}$, a series of Raman modes originating from two-phonon scattering processes involving combinations of optical modes of $\mathrm{E}$ and $\mathrm{A_1}$ symmetry with acoustic modes (TA, LA) can be distinguished.
	These include the following Raman peaks: $\mathrm{E}^5 + \mathrm{TA}$ (572~cm$^{-1}$), $\mathrm{A}^3_{1} + \mathrm{TA}$ (589~cm$^{-1}$), $\mathrm{E}^4 + \mathrm{LA}$ (602~cm$^{-1}$), $\mathrm{E}^5 + \mathrm{LA}$ (613~cm$^{-1}$), $\mathrm{A}^3_{1} + \mathrm{LA}$ (635~cm$^{-1}$), and $\mathrm{E}^2 + 2\mathrm{TA}$  (643~cm$^{-1}$).
	A three-phonon process arising from the coupling between the $\mathrm{A^5_{1}}$ mode and the 2LA mode is also observed at 454~cm$^{-1}$. 
	Notably, the $\mathrm{A^3_{1}}$–LA mode at 178~cm$^{-1}$ appears only at temperatures above 100~K, indicating its thermally activated character due to the bosonic nature of phonons~\cite{golasa2014resonant}.
	
	Overall, the comparison between the calculated phonon dispersion and the experimental Raman spectra summarized in Table~S1 of the SI demonstrates that, under resonant excitation, the vibrational response of bilayer 3R MoS$_2$ involves contributions from both zone-centre and zone-edge phonons.
	This behaviour underscores the crucial role of exciton-mediated phonon activation in enabling phonons across the Brillouin zone and governing their temperature-dependent evolution under resonant Raman scattering conditions.
	
	\begin{figure}[t]
		\centering
		\includegraphics[width=\linewidth]{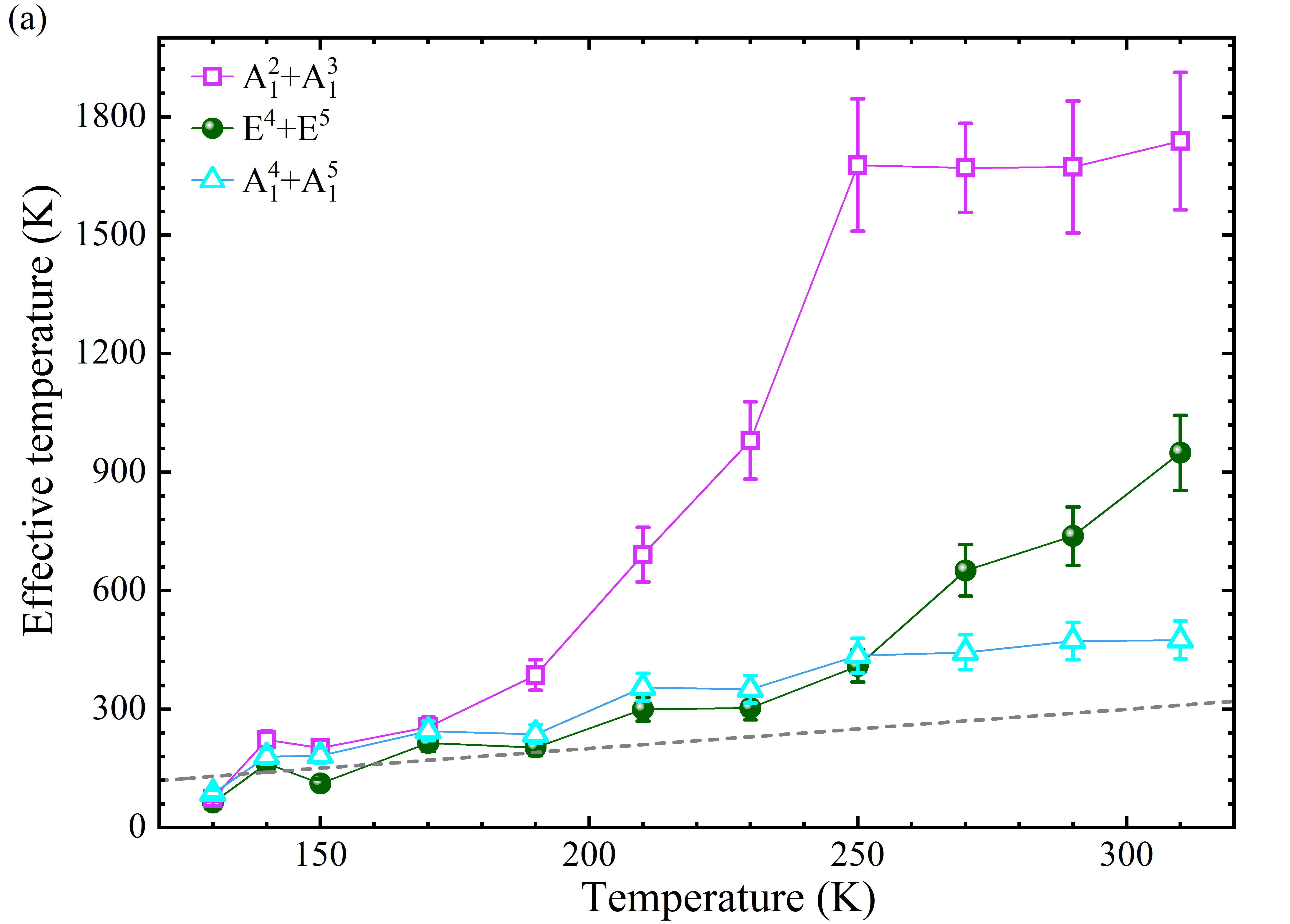}
		\caption{(a) Effective temperature extracted using the Boltzmann relation as a function of temperature for the $\mathrm{A^2_{1}}+\mathrm{A^3_{1}}$, $\mathrm{A^4_{1}}+\mathrm{A^5_{1}}$, and $\mathrm{E}^4+\mathrm{E}^5$ modes.The grey dashed line represents the condition $T_{eff}=T$, corresponding to the equality between the effective and measured temperatures.}
		\label{fig:fig4}
	\end{figure}
	
	\vspace{2\baselineskip}
	To elucidate the effect of resonant excitation conditions on the intensities of phonon modes in the Stokes and anti-Stokes branches, temperature-dependent RS measurements were performed under 1.96~eV excitation over the range 5--310~K.
	The integrated intensities were extracted by fitting the spectra with Lorentzian functions and a linear background, and the resulting temperature-dependent profiles for selected phonon modes are presented in Fig.~\ref{fig:fig3}.
	The temperature evolution reveals a clear and correlated behaviour of the Stokes and anti-Stokes components for the $\mathrm{A}^2_{1} + \mathrm{A}^3_{1}$, $\mathrm{A}^4_{1} + \mathrm{A}^5_{1}$, and $\mathrm{E}^4$ + $\mathrm{E}^5$ modes.
	The integrated intensities of closely spaced phonon modes $\mathrm{A}^2_{1} + \mathrm{A}^3_{1}$, $\mathrm{A}^4_{1} + \mathrm{A}^5_{1}$, and $\mathrm{E}^4$ + $\mathrm{E}^5$ were analysed as combined contributions, as the individual components could not be spectrally resolved (see Fig.~\ref{fig:fig2}(b)).
	For all the considered modes, the integrated intensities of the Stokes and anti-Stokes components exhibit similar temperature dependences. 
	First, the intensities decrease rapidly with increasing temperature from 5~K up to 120~K, by approximately a factor of 7 for the $\mathrm{A}^2_{1} + \mathrm{A}^3_{1}$ and $\mathrm{A}^4_{1}$ + $\mathrm{A}^5_{1}$ modes, and by nearly a factor of 20 for the $\mathrm{E}^4$ + $\mathrm{E}^5$ mode.
	This initial rapid quenching is followed by a nearly constant intensity up to 310~K.
	The integrated intensities of additional selected Raman peaks in the Stokes branch are analysed in detail in the SI.
	In particular, no measurable anti-Stokes signal is observed at low temperatures; it emerges only above 130~K and grows significantly with further temperature increase.
	This characteristic temperature marks the onset of a substantial phonon population contributing to the anti-Stokes scattering process.
	However, the coincidence between the temperature at which the Stokes-mode intensities change markedly and the onset of the anti-Stokes signal suggests that not only thermal activation of the phonon population is involved, but that resonant Raman scattering conditions must also be considered, as discussed in Figure~\ref{fig:fig4}.
	
	To further clarify the origin of this behaviour, we analyse the evolution of the Stokes and anti-Stokes scattering energies together with the $\mathrm{X_A}$ and $\mathrm{X_B}$ excitonic transitions, as shown in Figure~\ref{fig:fig4}(d)--(f). 
	The temperature dependence of the excitonic transitions continuously modifies the resonance conditions between the excitation energy $E_{\mathrm{L}} = 1.96$~eV and the phonon-shifted excitonic states.
	At low temperatures, the excitation energy lies close to the outgoing resonance condition involving the $\mathrm{X_A}$ exciton and the Stokes phonons.
	This resonance condition contributes to the strong enhancement of the Stokes Raman response and the high intensities observed in Figure~\ref{fig:fig4}(a)--(c).
	With increasing temperature, the $\mathrm{X_A}$ transition progressively redshifts away from the excitation energy, leading to a gradual detuning of the outgoing resonance condition and, consequently, to a pronounced quenching of the Stokes intensity up to approximately 120~K.
	A different behaviour is observed for the $\mathrm{X_B}$ exciton. 
	As the temperature increases, the anti-Stokes scattering energies progressively approach the incoming resonance condition with $\mathrm{X_B}$, while the Stokes branch simultaneously moves away from resonance.
	Around $T^{*} \approx 130$~K, where the anti-Stokes signal becomes experimentally detectable, the resonance conditions involving the $\mathrm{X_B}$ exciton begin to dominate the Raman response. In this temperature range, resonance effects originate from the proximity of the excitation energy to the $\mathrm{X_B}$ exciton and its phonon-shifted scattering channels.
	The opposite evolution of the Stokes and anti-Stokes branches indicates an asymmetric resonant Raman response. 
	The Stokes phonons are enhanced predominantly via outgoing resonance, when the excitation energy approaches the sum of the exciton and phonon energies, whereas the anti-Stokes phonons are governed mainly by incoming resonance, when the excitation energy approaches the excitonic transition energy directly.
	Our results therefore demonstrate that the interplay between different resonant RS mechanisms, driven by the temperature-induced tuning of the $\mathrm{X_A}$ and $\mathrm{X_B}$ excitonic transitions together with the thermal population of phonons, governs the evolution of the Raman response in bilayer 3R-MoS$_2$.
	
	\subsection*{Temperature-dependent Stokes and anti-Stokes phonon modes}
	
	Figure~\ref{fig:fig4}(a) presents the effective temperature ($T^{*}$) extracted from the intensity ratio of the Stokes phonon ($I_{\mathrm{S}}$) and the corresponding anti-Stokes component ($I_{\mathrm{AS}}$) using the Boltzmann relation, given by
	\begin{equation}
		\frac{I_{\mathrm{AS}}}{I_{\mathrm{S}}} = \exp\left(-\frac{\hbar \omega_{\mathrm{ph}}}{k T^{*}}\right),
	\end{equation}
	where $\omega_{\mathrm{ph}}$ is the phonon frequency, $\hbar$ is the reduced Planck constant, $k$ is the Boltzmann constant.
	At the lowest measured temperature (130~K), the effective temperature obtained for all modes closely follows the experimental temperature, indicating thermal equilibrium between the phonon population and the lattice. 
	However, above 170~K, a clear deviation emerges.
	In particular, the effective temperature associated with the $\mathrm{A}^2_{1}$ + $\mathrm{A}^3_{1}$ mode increases rapidly, reaching values as high as $T^{*} = 1800$~K at 310~K, whereas the $\mathrm{A}^4_{1}$ + $\mathrm{A}^5_{1}$ and $\mathrm{E}^4$ + $\mathrm{E}^5$ modes exhibit lower effective temperatures of approximately 900~K and 400~K, respectively.
	The pronounced deviation between the extracted effective and experimental temperatures demonstrates that the Stokes-to-anti-Stokes intensity ratio is strongly influenced by resonant Raman scattering conditions and therefore does not directly reflect the lattice temperature under resonant excitation., consistent with previous reports~\cite{maher2005resonance,maher2006temperature,goldstein2016raman}.

	\section*{Conclusion \label{Conclusion}}
	In summary, we have investigated the temperature-dependent resonant Raman response of bilayer 3R-MoS$_2$, combining RS, PL measurements, and theoretical modelling. 
	The evolution of excitonic transitions, described by the Varshni relation, enables continuous tuning of the resonance conditions between the A and B excitons over the 5–300~K temperature range.
	Under resonant excitation, the Raman spectra exhibit contributions from both zone-centre and finite-momentum phonons, including higher-order multiphonon processes activated by exciton–phonon coupling. 
	The temperature dependence of the Stokes and anti-Stokes intensities reveals a strong deviation from thermal equilibrium, reflected in the elevated effective phonon temperatures.
	We demonstrate that this behaviour originates from the interplay between incoming and outgoing resonant Raman scattering processes in the vicinity of the B exciton. The competition between these mechanisms, combined with the temperature-induced evolution of excitonic transitions and phonon populations, governs the enhancement of the Raman response.
	These findings provide new insight into exciton–phonon interactions in layered semiconductors and highlight the importance of resonant conditions in tailoring the vibrational properties of 2D materials.

	\section*{Methods \label{methods}}
	\subsection*{Sample fabrication}
	Bilayer $\mathrm{3R}\text{-}\mathrm{MoS}_{2}$ crystals were synthesized via atmospheric-pressure chemical vapor deposition (CVD). 
	A mixture of KI and $\mathrm{MoO}_{3}$ powders was placed in an alumina boat at the centre of a quartz tube furnace, with a $\mathrm{SiO}_{2}/\mathrm{Si}$ substrate positioned face-down above the precursor.
	Sulfur ($\mathrm{S}$) powder, used as the chalcogen source, was placed upstream and heated to $200\,^{\circ}\mathrm{C}$ using a heating tape to enable sublimation. 
	The furnace was ramped to $680\,^{\circ}\mathrm{C}$ and held for 15~min under a forming gas atmosphere ($\mathrm{Ar}:\mathrm{H}_{2} = 95\%:5\%$).
	The temperature was then increased to $720\,^{\circ}\mathrm{C}$ within 1~min, during which the carrier gas flow direction was reversed. 
	The flow was subsequently restored to the forward direction, and the system was maintained at $720\,^{\circ}\mathrm{C}$ for 10~min before being allowed to cool naturally to room temperature.
	
	\subsection*{Photoluminescence and Raman scattering spectroscopies}
	PL and RS measurements were performed using three excitation energies provided by diode lasers, $\lambda = 561$~nm (2.21~eV) and $\lambda = 514.5$~nm (2.41~eV), and a He--Ne laser, $\lambda = 632.8$~nm (1.96~eV).
	The laser beam, cleaned using laser-line or Bragg filters, was focused through a 50$\times$ long-working-distance objective with a numerical aperture of 0.55, producing a spot of approximately 1~$\mu$m in diameter.
	The signal was collected through the same objective, dispersed by a 0.75~m spectrometer, and detected using a liquid nitrogen-cooled charge-coupled device camera.
	Temperature-dependent PL and Raman measurements were carried out with the sample mounted on a cold finger in a continuous-flow cryostat equipped with x--y motorized positioners.
	The temperature was varied from 5 to 310~K, and the signal was collected after thermal stabilization.
	The excitation power at the sample was kept constant at 300~$\mu$W throughout all measurements to ensure a strong signal while avoiding local heating.
	
	\subsection*{Second harmonic generation}
	SHG measurements were performed using a pulsed femtosecond laser operating at 80~MHz with a fundamental wavelength of $\lambda = 785$~nm. 
	The excitation beam was passed through a linear polarizer and a $\lambda/2$ plate to control its polarization and focused onto the sample at normal incidence using a microscope objective. 
	The SHG signal was collected in reflection geometry, passed through a $\lambda/2$ plate and a linear polarizer for polarization-resolved detection, and detected using a liquid nitrogen-cooled CCD camera.
	
	\subsection*{First-principles calculations}
	DFT calculations were conducted in Vienna Ab initio Simulation Package~\cite{VASP} with Projector Augmented Wave method~\cite{PAW}. 
	Perdew–Burke–Ernzerhof parametrization~\cite{PBE} of general gradients approximation to the exchange-correlation functional was used. 
	The plane waves basis cutoff energy was set to 500 eV and a 12$\times$12$\times$1 $\Gamma$-centered Monkhorst-Pack k-grid sampling was applied. The geometric structure was optimized with $10^{-5}$ eV/\AA and 0.01 kbar criteria for the interatomic forces and stress tensor components, respectively. 
	Grimme's D3 correction was applied to describe the interlayer vdW interactions~\cite{D3}. 
	The phonon band structure of BL MoS$_2$ was calculated within Parli\'nski-Li-Kawazoe method~\cite{Parlinski}, as implemented in Phonopy software~\cite{Phonopy}. 
	A 3$\times$3$\times$1 supercells was found sufficient to converge the interatomic force constants within the harmonic approximation.

	\section*{Acknowledgments \label{Acknowledgements}}
	The work has been supported by the National Science Centre, Poland (grant no. 2023/50/O/ST3/00310).
	
	\section*{Author contributions \label{Author Contributions}}
	M.R.M. and Z.C. initiated and supervised the project.
	C.J. and Z.C. prepared the $\mathrm{3R}\text{-}\mathrm{MoS}_{2}$ samples.
	C.K.M. and K.W. performed the RS measurements.
	C.K.M. carried out the PL and SHG measurements.
	T.W. calculated the phonon dispersion using DFT.
	C.K.M., K.W., A.B., C.F., and M.R.M. analysed the experimental data.
	C.K.M. and M.R.M. wrote the manuscript with input from all co-authors.

	\section*{Conflicts of interest}
	There are no conflicts to declare.
	
	\section*{Availability of Data and Materials}
	The datasets generated and analysed during the current study are publicly available at the following link: XXX.
	
	\bibliographystyle{apsrev4-2}
	\bibliography{biblio}
	\onecolumngrid
\clearpage

\begin{center}
	
	{\Large \textbf{Supporting Information}}\\[1.0cm]
	
	{\large \textbf{Resonant Raman scattering in bilayer 3R-MoS$_2$}}\\[0.6cm]
	
	Chinmay K. Mohanty,$^{1,*}$
	Kacper Walczyk,$^{1}$
	Tomasz Wo\'zniak,$^{2}$
	Chengcheng Jiang,$^{3}$\\
	Adam Babi\'nski,$^{1}$
	Clement Faugeras,$^{4}$
	Zhaolong Chen,$^{3}$
	and Maciej R. Molas$^{1,\dagger}$\\[0.25cm]

	{\small \itshape
		$^{1}$Faculty of Physics, University of Warsaw, 02-093 Warsaw, Poland\\
		$^{2}$Institute of Theoretical Physics, Wroc{\l}aw University of Science and Technology, 50-370 Wroc{\l}aw, Poland\\
		$^{3}$Guangdong Provincial Key Laboratory of Nano-Micro Materials Research, School of Advanced Materials,\\
		Peking University Shenzhen Graduate School, Shenzhen, 518055, China\\
		$^{4}$LNCMI-EMFL, CNRS, UPR 3228, Universit\'e Grenoble Alpes, 38000 Grenoble, France
	}
	
	\end{center}
	
	\vspace{0.6cm}

\setcounter{section}{0}
\renewcommand{\thesection}{S\arabic{section}}

\setcounter{figure}{0}
\renewcommand{\thefigure}{S\arabic{figure}}

\section{Second harmonic generation studies of 3R-MoS$_2$}

	\begin{figure*}[h]
		\centering
		\includegraphics[width=.7\linewidth]{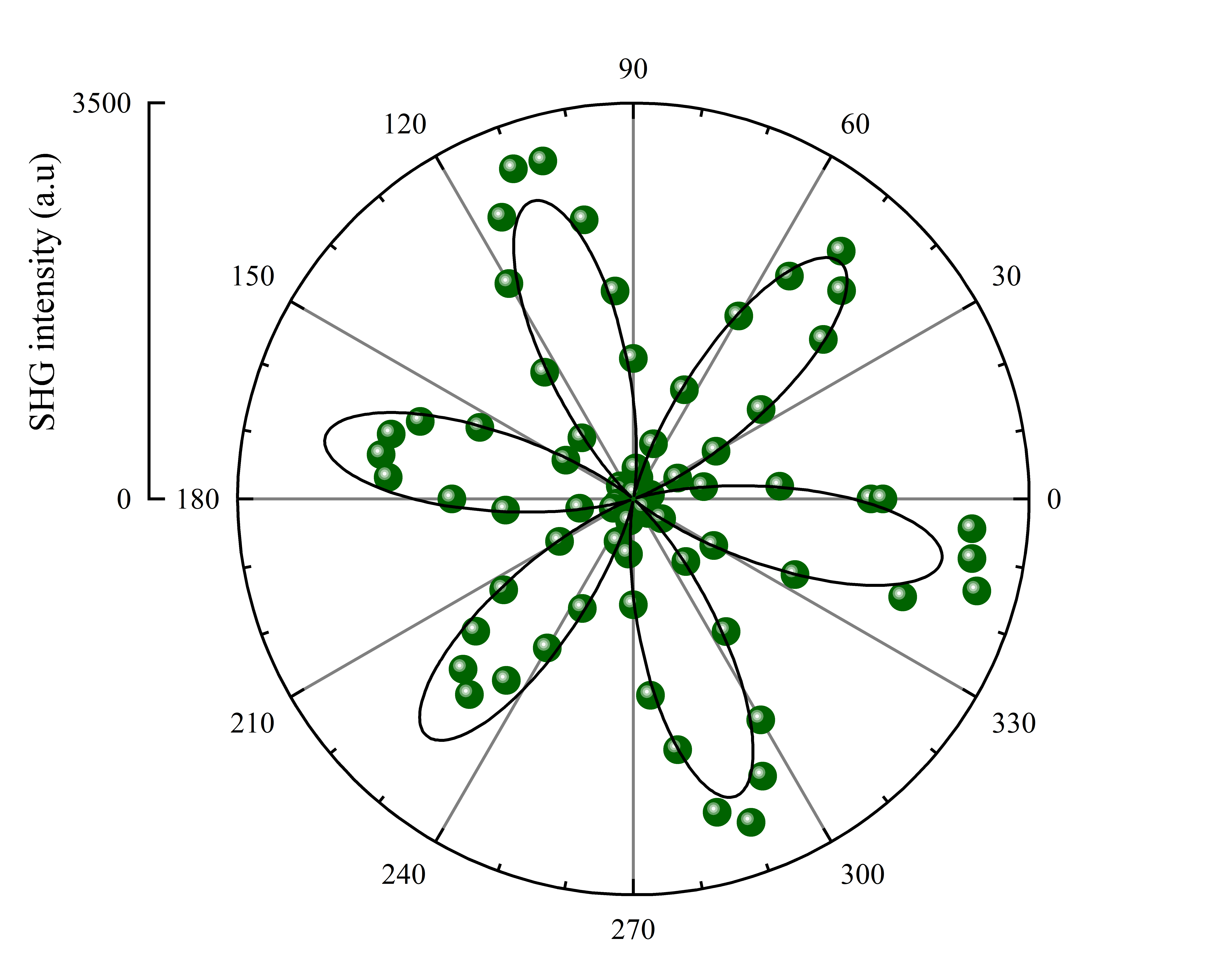}
		\caption{Angular dependence of the second-harmonic generation (SHG) intensity measured from a triangular monolayer 3R-stacked MoS$_2$ flake grown by chemical vapor deposition (CVD). The experimental data are shown as green circles, while the black solid line represents the fit based on the SHG model.  
		}
		\label{fig:s1}
	\end{figure*}
	
	Fig.~\ref{fig:s1} shows the second harmonic generation (SHG) intensity recorded as a function of the analyzer angle for the investigated MoS$_2$ bilayer (BL). A pronounced SHG signal is observed over the full angular range, indicating the non-centrosymmetric nature of the sample. The polar dependence of the SHG intensity exhibits a clear six-fold rotational symmetry, which is a characteristic signature of the 3R stacking configuration. 
	The angular dependence of the SHG intensity can be well described by the following relation:
	\begin{equation}
		I(\theta) = A\cos^{2}(3\theta + \phi) + B,
	\end{equation}
	where $A$ is the amplitude of the SHG signal, $\phi$ is the phase offset related to the crystal orientation, and $B$ accounts for the background contribution.
	This behaviour is consistent with the expected angular dependence for materials belonging to the $C_{3v}$ point group~\cite{Malard2013,Li2013}. In contrast, centrosymmetric structures such as 2H-stacked bilayers do not exhibit a strong SHG response due to inversion symmetry, which suppresses second-order nonlinear optical processes. The strong SHG response observed in the present sample therefore confirms the absence of inversion symmetry, thereby verifying the 3R stacking configuration of the investigated MoS$_2$ bilayer.
	
	\vspace{2em}
	\section{Calculated phonon density of states for BL 3R-MoS$_2$}
	\begin{figure*}[h!]
		\centering
		\includegraphics[width=0.7\linewidth]{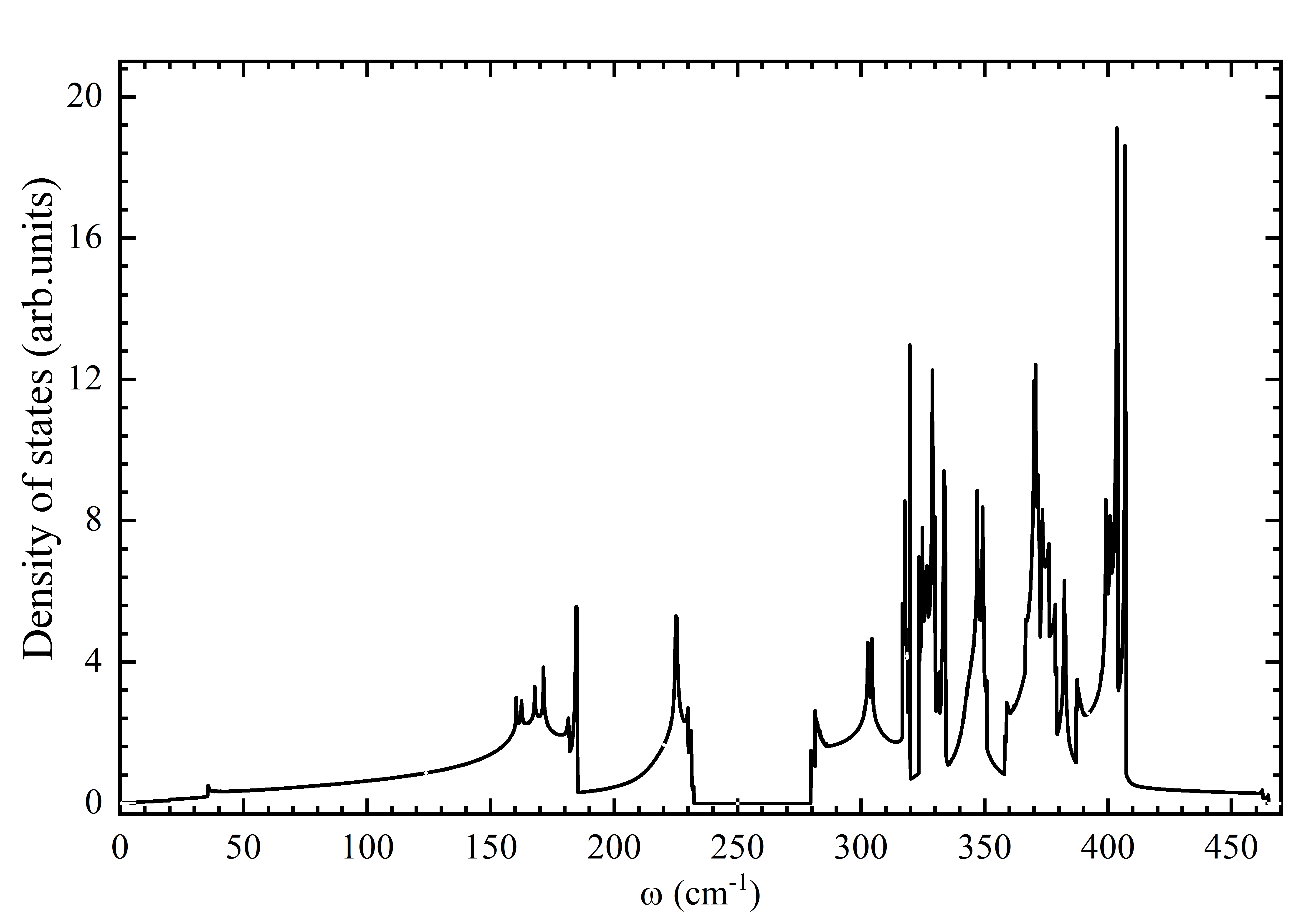}
		\caption{Phonon density of states for bilayer 2H-MoS2.}
		\label{fig:s2}
	\end{figure*}
	
	Fig.~\ref{fig:s2} presents the phonon density of states (PDOS) for bilayer MoS$_2$, calculated from first principles. The spectrum comprises contributions from both acoustic and optical phonon modes over the entire frequency range. The phonon states below $\sim 250$~cm$^{-1}$ are primarily associated with acoustic vibrations, whereas the higher-frequency region is dominated by optical phonon modes.
		
		\vspace{20em}
	\begin{table*}[h]
		\centering
		\caption{Comparison of experimentally measured and calculated Raman mode frequencies for bilayer MoS$_2$.}
		
			\vspace{3em}
		\label{tab:raman}
		
		\small
		
		\begin{tabular}{l @{\hspace{20pt}} c @{\hspace{20pt}} c}
			\hline\hline
			Mode & Measured (cm$^{-1}$) & Calculated (cm$^{-1}$) \\
			\hline
			TA(K) & 185 & 187 \\
			LA(K) & 232 & 226 \\
			$\mathrm{E}^2$ & 277 & 277 \\
			$\mathrm{E}^3$ & 279 & 279 \\
			$\mathrm{E}^4$ & 378 & 378 \\
			$\mathrm{E}^5$ & 382 & 383 \\
			$\mathrm{A}^2_1$ & 402 & 402 \\
			$\mathrm{A}^3_1$ & 404 & 404 \\
			$\mathrm{A}^4_1$ & 462 & 464 \\
			$\mathrm{A}^5_1$ & 465 & 466 \\
			LA+TA & 417 & 420 \\
			2LA & 464 & 454 \\
			$\mathrm{E}^5$+TA & 572 & 569 \\
			$\mathrm{E}^4$+LA & 602 & 600 \\
			$\mathrm{A}^3_1$+TA & 589 & 589 \\
			$\mathrm{E}^5$+LA & 611 & 613 \\
			$\mathrm{E}^2$+2TA & 643 & 643 \\
			$\mathrm{A}^3_1$-LA & 178 & 178 \\
			\hline\hline
		\end{tabular}
		
	\end{table*}
		
	\vspace{8em}
	\newpage
	\section{Temperature dependent Raman response of selected phonon modes under resonant excitation}

	\begin{figure*}[h!]
		\centering
		\includegraphics[width=1\linewidth]{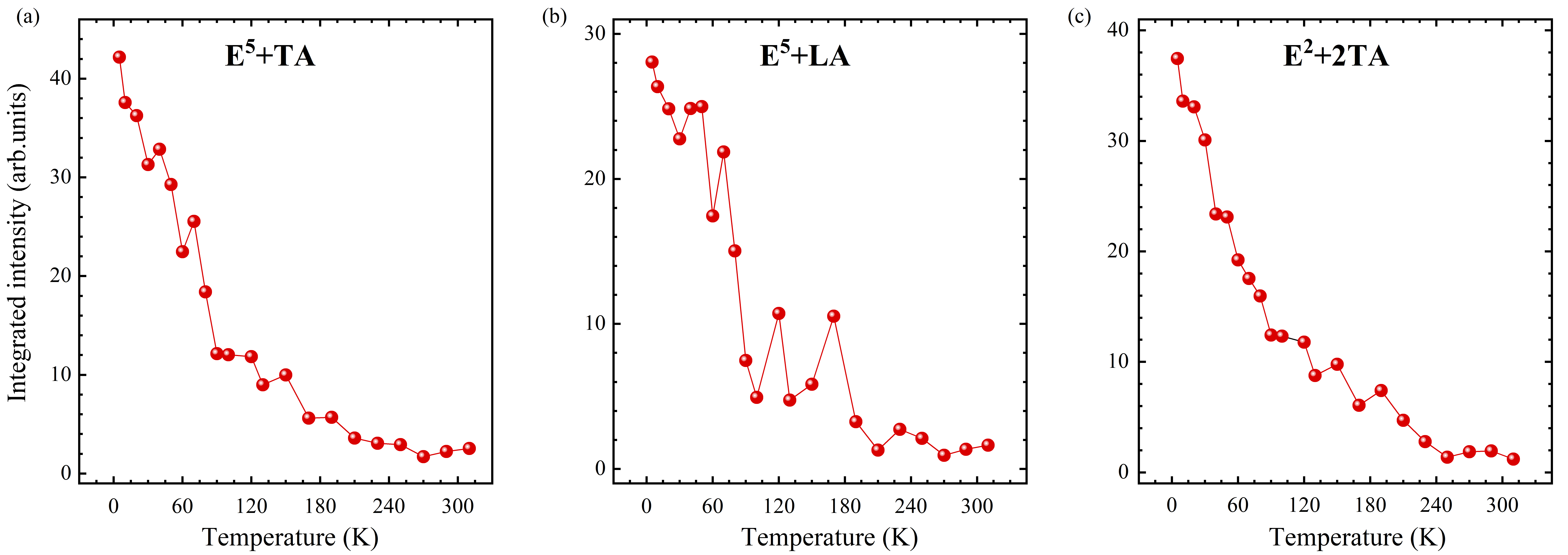}
		\caption{Temperature evolution of the integrated intensities of selected Raman modes: (a)$\mathrm{E}^5+\mathrm{TA}$, (b) $\mathrm{E}^5+\mathrm{LA}$, and (c) $\mathrm{E}^2+2\mathrm{TA}$. The data are extracted from temperature-dependent Raman spectra measured under resonant excitation. Note that the vertical axis ranges are adjusted to the maximum intensity of each mode.}
		\label{fig:s3}
	\end{figure*}
	\vspace{8em}
	Fig.~\ref{fig:s3} shows the evolution of the integrated intensities of selected Raman modes in the Stokes branch as a function of temperature.
	For all modes, the intensities decrease markedly from 5~K up to approximately 120~K, followed by a weak variation at higher temperatures. 
	This behaviour indicates a progressive quenching of the Raman response.
	No measurable anti-Stokes signal is detected for these phonon modes over the investigated temperature range, indicating that the scattering processes are dominated by the Stokes contribution.
	The intensity evolution is governed by temperature-dependent resonance conditions and exciton--phonon coupling. 
	A similar trend is found for the phonon modes discussed in the main text, indicating that this behaviour is a general feature of the resonant Raman response in the investigated system.
	
\end{document}